\documentstyle[12pt]{article}
\newcommand{\beq}{\begin{equation}}
\newcommand{\eeq}{\end{equation}}
\newcommand{\ba}{\begin{eqnarray}}
\newcommand{\ea}{\end{eqnarray}}

\newcommand{\dsl}
  {\kern.06em\hbox{\raise.15ex\hbox{$/$}\kern-.56em\hbox{$\partial$}}}

\newcommand{\eeqarr}{\end{eqnarray}}
\newcommand{\ZZ}{{\rm \kern 0.275em Z \kern -0.92em Z}\;}
\begin{document}
\begin{center}
{\Large Duality between Noncommutative Yang-Mills-Chern-Simons and
Non-Abelian Self-Dual Models}
\\
\vspace*{1.2cm}
{\large M. Botta Cantcheff$^{\dag,}$\footnote{botta@cbpf.br} and Pablo
Minces$^{\ddag,}$\footnote{pablo@fma.if.usp.br}}
\\
\vspace*{0.7cm}
$^{\dag}$Centro Brasileiro de Pesquisas F\'{\i}sicas (CBPF),\\
Departamento de Teoria de Campos e Part\'{\i}culas (DCP),\\ Rua Dr. Xavier
Sigaud 150, 22290-180, Rio de Janeiro, RJ, Brasil.\\
\vspace*{0.3cm}
$^{\ddag}$Instituto de F\'{\i}sica Te\'orica (IFT), Universidade Estadual
Paulista (UNESP),\\
Rua Pamplona 145, 01405-900, S\~ao Paulo, SP, Brasil.
\vspace*{0.5cm}
\end{center}
\begin{abstract}

By introducing an appropriate parent action and considering a perturbative
approach, we establish, up to fourth order terms in the field and for the
full range of the coupling constant, the equivalence between the
noncommutative Yang-Mills-Chern-Simons theory and
the noncommutative, non-Abelian Self-Dual model. In doing this, we 
consider two different approaches by using both the Moyal star-product 
and the Seiberg-Witten map.

\end{abstract}

\vskip 0.5cm

\section{Introduction}

The duality between the (2+1)-dimensional Maxwell-Chern-Simons (MCS)
\cite{deser} and Self-Dual (SD) \cite{townsend} Abelian models has been
established long time ago in \cite{deser2}, where a parent action for
both theories was introduced (see also \cite{banerjee2} for an 
approach introducing a master Lagrangian which has a gauge invariance 
in all fundamental fields).

In view of this, it is natural to investigate if such equivalence could be
extended to the non-Abelian case, by considering the
Yang-Mills-Chern-Simons (YMCS) theory and the non-Abelian SD model.
However, the non-Abelian situation is more involved, and the duality has
been established only
for the weak coupling regime \cite{bralic} (see also \cite{banerjee1} 
for an approach using hamiltonian techniques). In addition, it has been
argued in \cite{karlhede} that the use of a master action in the
non-Abelian situation is ineffective since YMCS and SD happen to be
dual to non-local theories each. The non-Abelian situation has also been
tackled in \cite{wot2} from a different point of view, by performing a
duality mapping which is realized by an iterative embedding of Noether
counterterms.

Recently, the parent action method has been brought back in \cite{bota} by
considering a novel perturbative analysis which, for the full range of
coupling constant, has established that the parent action actually
interpolates YMCS with a dual theory whose action is non-Abelian
Self-Dual model up to fourth order in the field, thus extending the proof
in \cite{deser2} to the non-Abelian case. In this formalism, the fourth
order terms are expected to be non-local.

The present paper deals with the noncommutative (NC) extension of
the duality between YMCS theory and non-Abelian SD model. In general, 
NC versions of the usual quantum field theories are 
obtained by replacing in all Lagrangians the usual product with the 
Groenewold-Moyal star-product \cite{groe}\cite{moyal} of the form

\beq
g(x)\star h(x)=
exp\left[\frac{i}{2}\;\theta^{\alpha\beta}\partial^{g}_{\alpha}
\partial^{h}_{\beta}\right]g(x)h(x)\; ,
\label{1} 
\eeq
where $g(x)$ and $h(x)$ are arbitrary functions and
$\theta^{\alpha\beta}$ is an antisymmetric constant tensor.

In recent years, NC field theories have generated a great deal of 
attention, due to the fact that they arise as low-energy descriptions 
of 
string backgrounds with antisymmetric tensor fields \cite{witten} (see
\cite{douglas}\cite{szabo} for reviews and additional references). In 
particular, an important result in \cite{witten} is that there exists 
a mapping, the Seiberg-Witten Map (SWM), which interpolates 
between a gauge theory and its NC counterpart in such a
way that gauge orbits are mapped into NC gauge orbits. The SWM is 
unique in the lowest non-trivial order of the NC parameter. 

The purpose of this paper is to investigate if the results regarding the
equivalence between YMCS theory and non-Abelian SD model can be extended
to the NC case. In order to do this, we will first introduce a suitable 
master action in which all usual products are replaced with the 
star-product, and consider the parent action method under a 
perturbative approach, as in \cite{bota}. In doing this, we will 
explicitly show that, in fact, the action for the NC YMCS theory is 
equivalent to that of the NC non-Abelian SD model, up to fourth 
order in the field, and for the full range of the coupling constant.

Then, we will consider an alternative approach to the problem, 
involving to perform, to the first non-trivial order in
the NC parameter, the SWM to the usual 
commutative master action. 
By considering again a perturbative approach, as in \cite{bota}, we 
will show that the known equivalence between YMCS theory and 
non-Abelian SD model, up to fourth order terms in the field and for the 
full 
range of the coupling constant, is maintained in the NC space. This result is 
not trivial also because commutative non-Abelian SD and YMCS theories generalize 
to 
NC field theories in different ways, due to the fact that the YMCS 
theory is gauge invariant, whereas the non-abelian SD model is not. It is knowm 
that, whereas 
gauge theories 
are lifted to their NC couterparts via the SWM, non-gauge theories are affected 
only in the products of the fields in the Lagrangian. In fact, our results 
could be considered as the extension to the non-Abelian case of the results in 
\cite{ghosh} regarding the survival of the equivalence between the SD and MCS 
Abelian models when the space-time becomes NC.

The paper is organized as follows. In Section 2 we deal with a parent 
action in which all usual products are replaced with star-products, 
and, by 
considering a perturbative approach, we establish, up to fourth order 
terms in the field and for the full range of the coupling constant, the 
equivalence between YMCS and non-Abelian SD models in which the usual 
products are replaced with star-products. In Section 3 we perform a 
SWM to the usual parent action and, by considering again a perturbative 
approach, we show that the duality between commutative YMCS and 
non-Abelian SD models survives when the space-time becomes NC.

\section{The Star-Product}

We begin by proposing a parent action in which all usual products are 
replaced with star-products. It is given by

\beq I_{P}=\int d^{3}x\; Tr\left[-\mu\; f^{\mu} \star f_{\mu} +
m\;\epsilon^{\alpha\mu\nu}\left(f_{\alpha}\star
F_{\mu\nu}-A_{\alpha}\star \left(F_{\mu\nu}-\frac{2}{3}\;
A_{\mu}\star A_{\nu}\right)\right)\right]\; , \label{2} \eeq where
$\mu$ and $m$ are, in principle, arbitrary coefficients, our
metric convention is $\eta_{\mu\nu} =diag(-,+,+)$, and \beq
F_{\mu\nu}=\partial_{\mu}A_{\nu}-\partial_{\nu}A_{\mu}+\left[A_{\mu}
\;,\; A_{\nu}\right]_{M}\; , \label{3} \eeq \beq
\left[A_{\mu}\;,\; A_{\nu}\right]_{M}=A_{\mu}\star
A_{\nu}-A_{\nu}\star A_{\mu}\; . \label{4} \eeq Here
$A_{\mu}=A^{a}_{\mu}\tau^{a}$ and $f_{\mu}=f^{a}_{\mu}\tau^{a}$ are
fields in the adjoint
representation of a non-Abelian gauge group $G$ (with $a=1\cdots
dim\; G$), and $\tau^{a}$ are the matrices representing $G$, such
as $[\tau^{a},\tau^{b}]=\tau^{abc}\tau^{c}$, where $\tau^{abc}$
are the structure constants of the group.

We shall verify that solving $I_P$, first for $f_{\mu}$ (in
terms of $A_{\mu}$) and further for $A_{\mu}$ (in terms of $f_{\mu}$) we
recover both the NC YMCS theory and the NC non-Abelian Self-Dual model,
respectively.

Integration by parts in Eq.(\ref{1}) yields

\ba
g(x)\star h(x) &=& g(x)h(x) + \cdots\; ,\nonumber\\
g(x)\star h(x)\star p(x) &=& g(x)\left(h(x)\star p(x)\right) +\cdots\; ,
\label{5}
\ea
where the dots stand for total derivatives. Throughout this paper, we
consider boundary conditions such as the surface terms in the action
vanish. Then, using Eqs.(\ref{5}) we can write $I_{P}$ as

\beq I_{P}=\int d^{3}x\; Tr\left[-\mu\; f^{\mu}f_{\mu} +
m\;\epsilon^{\alpha\mu\nu}\left(f_{\alpha}F_{\mu\nu}-A_{\alpha}
\left(F_{\mu\nu}-\frac{2}{3}\; A_{\mu}\star
A_{\nu}\right)\right)\right]\; . \label{6} \eeq From the above
expression we get the following equation of motion for $f_{\mu}$

\beq f^{\mu}=\frac{m}{2\mu}\;\epsilon^{\alpha\mu\nu}F_{\mu\nu}\; .
\label{7} \eeq Introducing this result into Eq.(\ref{6}) and using
again Eqs.(5) we get

\beq I_{P}=\int d^{3}x\; Tr\left[-\frac{m^{2}}{2\mu}\;
F^{\mu\nu}\star F_{\mu\nu} -
m\;\epsilon^{\alpha\mu\nu}A_{\alpha}\star
\left(F_{\mu\nu}-\frac{2}{3}\; A_{\mu}\star
A_{\nu}\right)\right]\; , \label{8} \eeq which is the action for
the NC YMCS theory. Notice that the single parameter which
effectively appears in the
 theory is the boson mass, $M \equiv  \frac{- 2 \mu }{m}\;$.

Now we look for the Self-Dual model. Note that we can write $I_{P}$ as
follows

\beq I_{P}=\int d^{3}x\; Tr\left[-\mu\; f^{\mu}\star f_{\mu} +
2m\;\epsilon^{\alpha\mu\nu}\left(\left(f_{\alpha}-A_{\alpha}\right)\star\partial_{\mu}A_{\nu}
+\left(f_{\alpha}-\frac{2}{3}\;A_{\alpha}\right)\star A_{\mu}\star
A_{\nu}\right)\right]\; , \label{99} \eeq and using Eq.(\ref{5})
we get

\beq I_{P}=\int d^{3}x\; Tr\left[-\mu\; f^{\mu}f_{\mu} +
2m\;\epsilon^{\alpha\mu\nu}\left(\left(f_{\alpha}-A_{\alpha}\right)\partial_{\mu}A_{\nu}
+\left(\left(f_{\alpha}-\frac{2}{3}\;A_{\alpha}\right)\star
A_{\mu}\right)A_{\nu}\right)\right]\; . \label{9} \eeq From the
above expression, we find the following equation of motion for
$A_{\mu}$

\beq
0=\epsilon^{\alpha\mu\nu}\left(\partial_{\mu}f_{\nu}-2\;\partial_{\mu}A_{\nu}+\left[A_{\mu}\;
,\;f_{\nu}\right]_{M}-\left[A_{\mu}\; ,\;
A_{\nu}\right]_{M}\right)\; , \label{10} \eeq which after
contracting with a Levi-Civita tensor gives rise to \beq
2F_{\mu\nu}=\partial_{\mu}f_{\nu}-\partial_{\nu}f_{\mu}
+\left[A_{\mu}\; ,\;f_{\nu}\right]_{M}-\left[A_{\nu}\;
,\;f_{\mu}\right]_{M}\; . \label{11} \eeq Now we must find a
solution $A_{\mu}=A_{\mu}[f_{\nu}]$ which should be
replaced into Eq.(\ref{99}), thus getting an action which is a
functional of $f_{\mu}$. However, as in the commutative
case, the problem regarding the equation above is that it cannot
be inverted.

In order to deal with this problem, we assume that a solution exists
at least perturbatively, and look for a solution of the form

\beq
A_{\mu}=A_{\mu}^{(0)}+A_{\mu}^{(1)}\left[f_{\nu}\right]+A_{\mu}^{(2)}\left[f_{\nu}\right]+\cdots
\; . \label{12} \eeq Then, the lowest order in Eq.(\ref{11})
yields

\beq
F^{(0)}_{\mu\nu}=0\; ,
\label{13}
\eeq
so that $A_{\mu}^{(0)}$ is just a pure-gauge, and we will not consider it
anymore. The following order in Eq.(\ref{11}) gives

\ba 0 &=&
\partial_{\mu}\left(2A^{(1)}_{\nu}-f_{\nu}\right)-\partial_{\nu}\left(2A^{(1)}_{\mu}-f_{\mu}\right) + \left[A^{(0)}_{\mu}\; ,\; 2A^{(1)}_{\nu}-f_{\nu}\right]_{M} \nonumber\\
&-&\left[A^{(0)}_{\nu}\; ,\; 2A^{(1)}_{\mu}- f_{\mu}\right]_{M}\;
, \label{14} \ea which has solution

\beq A^{(1)}_{\mu}=\frac{1}{2}\; f_{\mu}\; . \label{15} \eeq Now
we will show that in fact the explicit expression of
$A_{\mu}^{(2)}$ is not needed for our present purposes.
Introducing Eq.(\ref{12}) into Eq.(\ref{99}), using Eq.(\ref{5})
and discarding the pure-gauge term $A_{\mu}^{(0)}$ we find

\beq I_{P}=I_{P}\left[A^{(1)}_{\mu}\right]+ \int d^{3}x\;
Tr\left[m\epsilon^{\alpha\mu\nu}
\left(f_{\nu}-2A^{(1)}_{\nu}\right)
\left(\partial_{\alpha}A^{(2)}_{\mu}-\partial_{\mu}A^{(2)}_{\alpha}\right)
\right]+{\cal O}(f^{4})\; . \label{16} \eeq Then, using
Eq.(\ref{15}) we get

\beq I_{P}=I_{P}\left[A^{(1)}_{\mu}\right]+{\cal O}(f^{4})\; .
\label{17} \eeq From Eqs.(\ref{99}, \ref{15}, \ref{17}) we finally
get

\beq I_{P}=\int d^{3}x\; Tr\left[-\mu\; f^{\mu} \star f_{\mu} +
\frac{m}{2}\;\epsilon^{\alpha\mu\nu}f_{\alpha}\star
\left(\partial_{\mu}f_{\nu}+\frac{2}{3}\; f_{\mu}\star
f_{\nu}\right)\right] +{\cal O}(f^{4})\; , \label{18} \eeq where
we have reproduced the action for the NC non-Abelian SD model. In
this theory we also find $M \equiv  \frac{- 2 \mu }{m}$ as the
single arbitrary constant which characterizes the model. From
Eqs.(\ref{8}, \ref{18}) we see that, as anticipated, we have shown
that NC YMCS is dual to a theory which coincides with the NC
non-Abelian SD model, up to fourth order in $f_{\mu}$, and for the
full range of the coupling constant.

In the following section, we will tackle this problem under a different 
point of view, namely, that of the SWM, obtaining analogous results.

\section{Seiberg-Witten Map and Duality}

Our starting point is the following parent action in the usual 
commutative space-time

\beq
S_{P}=\int d^{3}x\; Tr\left[-\mu\; f^{\mu} f_{\mu} +
2m\;\epsilon^{\alpha\mu\nu}\left(f_{\alpha}
\left(\partial_{\mu}A_{\nu}+A_{\mu}A_{\nu}\right)-
A_{\alpha}\left(\partial_{\mu}A_{\nu}+\frac{2}{3}\; 
A_{\mu}A_{\nu}\right)\right)\right]\; ,
\label{500}
\eeq
where $A_{\mu}$ and $f_{\mu}$ are fields in the adjoint representation 
of a non-Abelian gauge group. 

In this section, we will be concerned with the inverse SWM at the first 
non-trivial order in the NC parameter. It is given by

\beq
A_{\mu}=A'_{\mu}+a'_{\mu}(A'_{\nu}, \theta)\; ,
\label{501}
\eeq 
where

\beq
a'_{\mu}=\frac{1}{4}\;\theta^{\rho\sigma}\left[A'_{\rho}\; 
,\; 2\partial_{\sigma}
A'_{\mu}-\partial_{\mu}A'_{\sigma}+A'_{\sigma}A'_{\mu}-A'_{\mu}A'_{\sigma}
\right]_{+}\; .
\label{502}
\eeq
We emphasize that all calculations in this section will be valid up to 
$O(\theta^{2})$ terms. The above mapping together with $f_{\mu}=f'_{\mu}$ lifts 
$S_{P}$ to the following NC parent action

\ba
S'_{P}&=&\int d^{3}x\; Tr [-\mu\; f'^{\mu} f'_{\mu} +
2m\;\epsilon^{\alpha\mu\nu}\left(f'_{\alpha}-A'_{\alpha}-a'_{\alpha}\right)
\partial_{\mu}\left(A'_{\nu}+a'_{\nu}\right)\nonumber\\ &&
\qquad\qquad\quad +2m\;\epsilon^{\alpha\mu\nu}
\left(f'_{\alpha}-
\frac{2}{3}\;
A'_{\alpha}-\frac{2}{3}\;a'_{\alpha}\right)
\left(A'_{\mu}+a'_{\mu}\right)\left(A'_{\nu}+a'_{\nu}\right)]\; ,
\label{503}
\ea
which can be written as

\ba
S'_{P}&=&\int d^{3}x\; Tr [-\mu\; f'^{\mu} f'_{\mu} +
2m\;\epsilon^{\alpha\mu\nu}\left(f'_{\alpha}-A'_{\alpha}\right)
\partial_{\mu}\left(A'_{\nu}+a'_{\nu}\right)\nonumber\\ &&
\qquad\qquad\quad +2m\;\epsilon^{\alpha\mu\nu}
\left(f'_{\alpha}-
\frac{2}{3}\; 
A'_{\alpha}\right)\left(A'_{\mu}A'_{\nu}+a'_{\mu}A'_{\nu}
+A'_{\mu}a'_{\nu}\right)\nonumber\\ &&
\qquad\qquad\quad -2m\;\epsilon^{\alpha\mu\nu}
a'_{\alpha}\left(\partial_{\mu}A'_{\nu}+
\frac{2}{3}\; A'_{\mu}A'_{\nu}\right)]\; +\; O(\theta^{2}).
\label{504}
\ea
We will show that solving $S'_{P}$, first for $f'^{\mu}$ and further 
for $A'^{\mu}$, we recover the SWM-lifted actions for the YMCS theory 
and the non-Abelian SD model respectively, even when both theories generalize to 
NC field theories in different ways.

First, from Eq.(\ref{504}) we get the following equation of motion for 
$f'^{\mu}$

\beq
f'^{\mu}=\frac{m}{\mu}\;\epsilon^{\mu\alpha\beta}\left[
\partial_{\mu}\left(A'_{\nu}+a'_{\nu}\right)+A'_{\mu}A'_{\nu}+
a'_{\mu}A'_{\nu}+A'_{\mu}a'_{\nu}\right]\; +\; O(\theta^{2}),
\label{505}
\eeq
and introducing this back into Eq.(\ref{504}) we find

\ba 
S'_{P}&=&\int d^{3}x\; Tr [-\frac{m^{2}}{\mu}\; 
\left(\partial_{\mu}A'_{\nu}-\partial_{\nu}A'_{\mu}+A'_{\mu}A'_{\nu}
-A'_{\nu}A'_{\mu}\right)\nonumber\\ &&\qquad\qquad\qquad\quad\times
\left(\partial^{\mu}
\left(A'^{\nu}+a'^{\nu}\right)+A'^{\mu}A'^{\nu}+a'^{\mu}A'^{\nu}
+A'^{\mu}a'^{\nu}\right)\nonumber\\&&\qquad\qquad
-\frac{m^{2}}{\mu}\;\left(\partial_{\mu}a'_{\nu}-\partial_{\nu}a'_{\mu}
+a'_{\mu}A'_{\nu}-A'_{\nu}a'_{\mu}+A'_{\mu}a'_{\nu}-a'_{\nu}A'_{\mu}
\right)\nonumber\\ &&\qquad\qquad\qquad\quad\times
\left(\partial^{\mu}A'^{\nu}+A'^{\mu}A'^{\nu}\right)\nonumber
\\&&\qquad\qquad
-2m\;\epsilon^{\alpha\mu\nu}A'_{\alpha}\left(\partial_{\mu}
\left(A'_{\nu}+a'_{\nu}\right)+\frac{2}{3}\; 
\left(A'_{\mu}A'_{\nu}+a'_{\mu}A'_{\nu}+A'_{\mu}a'_{\nu}\right)\right)
\nonumber\\&&\qquad\qquad
-2m\;\epsilon^{\alpha\mu\nu}a'_{\alpha}\left(\partial_{\mu}A'_{\nu}
+\frac{2}{3}\;A'_{\mu}A'_{\nu}\right)]\; +\; O(\theta^{2}),
\label{506}
\ea
which to $O(\theta)$ is the SWM-lifted action of the YMCS theory

\ba
S'_{YMCS}&=&\int d^{3}x\; Tr 
[-\frac{m^{2}}{\mu}\;\left(\partial_{\mu}
\left(A'_{\nu}+a'_{\nu}\right)-\partial_{\nu}
\left(A'_{\mu}+a'_{\mu}\right)\right)\nonumber\\ 
&&\qquad\qquad\qquad\qquad\times
\left(\partial^{\mu}\left(A'^{\nu}+a'^{\nu}\right)+\left(A'^{\mu}+a'^{\mu}
\right)\left(A'^{\nu}+a'^{\nu}\right)\right)
\nonumber\\ &&\qquad\qquad
-\frac{m^{2}}{\mu}\;\left(\left(A'_{\mu}+a'_{\mu}\right)
\left(A'_{\nu}+a'_{\nu}\right)-\left(A'_{\nu}+a'_{\nu}\right)
\left(A'_{\mu}+a'_{\mu}\right)\right)
\nonumber\\
&&\qquad\qquad\qquad\qquad\times
\left(\partial^{\mu}\left(A'^{\nu}+a'^{\nu}\right)+\left(A'^{\mu}+a'^{\mu}
\right)\left(A'^{\nu}+a'^{\nu}\right)\right)
\nonumber\\ &&\qquad\qquad-2m\;\epsilon^{\alpha\mu\nu}
\left(A'_{\alpha}+a'_{\alpha}\right)
\nonumber\\
&&\qquad\qquad\qquad\qquad\times
\left(\partial_{\mu}
\left(A'_{\nu}+a'_{\nu}\right)+\frac{2}{3}\;\left(A'_{\mu}+a'_{\mu}
\right)\left(A'_{\nu}+a'_{\nu}\right)\right)]\; .\nonumber\\
\label{507}
\ea

Now we look for the SWM-lifted action of the SD model. From 
Eq.(\ref{504}), we find the following equation of motion for $A'_{\mu}$

\ba
0 &=& 
\epsilon^{\alpha\mu\nu}(\partial_{\mu}f'_{\nu}-2\partial_{\mu}
\left(A'_{\nu}+a'_{\nu}\right)+\left[A'_{\mu}+a'_{\mu}\; 
,\; f'_{\nu}\right]-\left[A'_{\mu}\; ,\;A'_{\nu}\right]
-\left[a'_{\mu}\; ,\;A'_{\nu}\right]\nonumber\\ &&\qquad 
-\left[A'_{\mu}\; 
,\;a'_{\nu}\right])+ O(\theta^{2})\; ,
\label{508}
\ea
and contracting with a Levi-Civita tensor we get

\ba
2\left(\partial_{\mu}\left(A'_{\nu}+a'_{\nu}\right)-\partial_{\nu}
\left(A'_{\mu}+a'_{\mu}\right)+\left[A'_{\mu}\; 
,\;A'_{\nu}\right]+\left[a'_{\mu}\; ,\;A'_{\nu}\right]
+\left[A'_{\mu}\; ,\;a'_{\nu}\right]\right)\nonumber\\
=\partial_{\mu}f'_{\nu}-\partial_{\nu}f'_{\mu}+
\left[A'_{\mu}+a'_{\mu}\; ,\; f'_{\nu}\right]-
\left[A'_{\nu}+a'_{\nu}\; ,\; f'_{\mu}\right] + O(\theta^{2})\; .
\label{509}
\ea

Now we must look for a solution $A'_{\mu}=A'_{\mu}[f'_{\nu}]$. 
However, the equation above cannot be inverted. To deal with this 
problem, we proceed as in the case of the previous section, and assume 
that a solution exists at least perturbatively, namely

\beq
A'_{\mu}=A^{'(0)}_{\mu}+A^{'(1)}_{\mu}\left[f'_{\nu}\right]+
A^{'(2)}_{\mu}\left[f'_{\nu}\right]+\cdots\; . 
\label{510} 
\eeq
All there is to do now is to solve Eq.(\ref{509}) order by order, 
as in the previous section. However, before doing so, we note that in 
fact Eq.(\ref{509}) differs from the following equation

\ba
2\left(\partial_{\mu}\left(A'_{\nu}+a'_{\nu}\right)-\partial_{\nu} 
\left(A'_{\mu}+a'_{\mu}\right)+\left[A'_{\mu}+a'_{\mu}\;
,\;A'_{\nu}+a'_{\nu}\right]\right)\nonumber\\
=\partial_{\mu}f'_{\nu}-\partial_{\nu}f'_{\mu}+
\left[A'_{\mu}+a'_{\mu}\; ,\; f'_{\nu}\right]-
\left[A'_{\nu}+a'_{\nu}\; ,\; f'_{\mu}\right]\; ,
\label{511}
\ea
only by $O(\theta^{2})$ terms. The interesting advantage regarding 
Eq.(\ref{511}) is that it will easily allow us to find the explicit 
expressions of $A^{'(0)}_{\mu}+a^{'(0)}_{\mu}$ and 
$A^{'(1)}_{\mu}+a^{'(1)}_{\mu}$ (up to $O(\theta^{2})$ terms), if not 
of $A^{'(0)}_{\mu}$ and $A^{'(1)}_{\mu}$. For our present purposes, it 
will suffice. Note that, in fact, the above equation has the same 
formal structure as Eq.(\ref{11}), and thus we easily find the solution

\beq
A^{'(1)}_{\mu}+a^{'(1)}_{\mu}=\frac{1}{2}\;f'_{\mu}+O(\theta^{2})\; ,
\label{512}
\eeq
whereas $A^{'(0)}_{\mu}+a^{'(0)}_{\mu}$ remains as a pure-gauge (also 
up to $O(\theta^{2})$ terms). As in the case of the previous section, 
it can be shown that $A^{'(2)}_{\mu}+a^{'(2)}_{\mu}$ only 
contributes to $O(f^{'4})$ and so we do not need to compute it. In fact 
it can be shown that, up to $O(\theta^{2})$ terms, the following 
identity stands

\beq
S'_{P}=S'_{P}\left[A^{'(1)}_{\mu}+a^{'(1)}_{\mu}\right] + O(f^{'4})\; .
\label{513}
\eeq
From Eqs.(\ref{503}, \ref{512}, \ref{513}) we find (up to 
$O(\theta^{2})$ terms)

\beq
S'_{P}=S'_{SD}+O(f'^{4})\; ,
\label{514}
\eeq
where $S'_{SD}$ is the action of the SWM-lifted non-Abelian SD model

\beq
S'_{SD}=\int d^{3}x\; Tr\left[-\mu\; f'^{\mu} f'_{\mu} +
\frac{m}{2}\;\epsilon^{\alpha\mu\nu}f'_{\alpha}
\left(\partial_{\mu}f'_{\nu}+\frac{2}{3}\; 
f'_{\mu}f'_{\nu}\right)\right]\; .
\label{515}
\eeq
In this way, we have shown that the duality between commutative YMCS 
theory and non-Abelian SD model, up to fourth order terms in the field, 
survives when the space-time becomes NC, even when both theories generalize
to NC field theories in different ways. We point out that it would be
interesting to extend our results to higher orders in $\theta$. 

One last comment concerns the relation between our approach and the 
results included in the previous literature regarding NC Chern-Simons 
theories. It is already known 
\cite{grandi}\cite{moreno1}\cite{moreno2} that the SWM connects the 
commutative and NC versions of two related models, namely, the 
Chern-Simons theory and the Wess-Zumino-Witten model. We may wonder 
if similar results could be found in the present case. In this respect, it 
should be mentioned that, in fact, the SWM lifts the NC Yang-Mills action
(defined through the usual Moyal product) into a non-polynomial 
commutative 
action which differs from the usual one. So, in principle, we should not 
expect the SWM to map the NC YMCS action Eq.(\ref{8}) into its commutative 
version, due to the presence of the Yang-Mills term. This 
is to be contrasted with our result Eq.(\ref{507}) which, as pointed out 
before, corresponds to the SWM-lifted action of the YMCS theory.

We hope that the novel perturbative method considered in this article 
could be helpful to establish other dual equivalences between models, and 
also to simplify the treatment of NC non-Abelian mathematical structures.

\section{Acknowledgements}

We would like to thank the referee for the useful and constructive 
suggestions. M.B.C. would also like to thank G. Barbosa for a critical 
reading of the manuscript. P.M. is deeply indebted to N. Berkovits for 
suggesting the topic of the present paper, and is also grateful to A. Yu. 
Petrov and C. Wotzasek for discussions. M.B.C. acknowledges financial 
support by CLAF. P.M. was supported by FAPESP grant 01/05770-1.

\end{document}